\documentclass[11pt,nofootinbib,floatfix]{revtex4}
\usepackage{mathrsfs}
\usepackage{graphicx}
\usepackage{amsmath}
\usepackage{amsfonts}
\usepackage{amssymb}
\usepackage{color}

\usepackage{epsfig}
\usepackage{CJK}
\usepackage{eepic}
\usepackage{bbm}
\usepackage{dcolumn}% Align table columns on decimal point
\usepackage{bm}% bold math
\usepackage{ulem}
\usepackage{slashbox}
\usepackage{multirow}% for the tabular, added in 30.01.2013

\newcommand{\omits}[1]{}

\def\bc{\begin{center}}
\def\nno{\nonumber}
\def\ec{\end{center}}
\def\be{\begin{eqnarray}}
\def\ee{\end{eqnarray}}

\definecolor{dyellow}{rgb}{1.,0.8,.0}
\definecolor{myblue}{rgb}{.1,.1,.7}
\definecolor{dcyan}{rgb}{.0,.6,.6}
\definecolor{cyan}{rgb}{0.4,1.0,1.0}
\definecolor{dmagenta}{rgb}{0.6,0.0,0.6}
\definecolor{brown}{rgb}{0.6,0.2,0.}
\definecolor{darkblue}{rgb}{.0,.0,0.5}
\definecolor{darkred}{rgb}{0.75,0.0,0.0}
\definecolor{orange}{rgb}{1.,.6,.0}
\definecolor{dorange}{rgb}{0.8,.4,.0}
\definecolor{green}{rgb}{0.0,1.0,0.0}
\definecolor{darkgreen}{rgb}{0.0,0.6,0.0}
\definecolor{purple}{rgb}{.4,.0,.4}
\definecolor{lightgrey}{rgb}{0.7, 0.7, 0.7}
\definecolor{grey}{rgb}{0.4, 0.4, 0.4}
\def\be{\begin{equation}}
\def\ee{\end{equation}}
\def\bea{\begin{eqnarray}}
\def\eea{\end{eqnarray}}
\def\e{\epsilon}

\def\o{\omega}

\def\p{\partial}

\def\a{\alpha}

\def\>{\rangle} %right angle
\def\<{\langle} %left angle

\def\der#1#2{{\p {#1} \over \p {#2}}}
\def\dder#1#2{{d {#1} \over d {#2}}}

\begin{document}

%\preprint{hep-th/yymmnnn}

\title{Distinguishing Schwinger effect from Hawking radiation in Reissner-Nordstr{\"o}m black holes via entanglement}

\author{Ruo-Han Wang$^{1}$} \email{wangrh67@mail2.sysu.edu.cn}
\author{Jia-Rui Sun$^{1}$} \email{sunjiarui@sysu.edu.cn}

\affiliation{${}^1$School of Physics and Astronomy, Sun Yat-Sen University, Guangzhou 510275, China}

%\date{May, 2014}

%% REVTEX4
%\maketitle

\begin{abstract}
A charged black hole can emit charged particles via two independent mechanisms: the Hawking radiation and the Schwinger effect, which are intertwined in the radiation spectrum. In this paper, we will show that the two effects can be distinguished by analyzing the entanglement entropy carried by the produced particle pairs. Explicitly, we apply the island formula to the near extremal Reissner-Nordstr{\"o}m (RN) black hole to calculate the total entanglement entropy of the radiation. Meanwhile we use the heat kernel method to calculate the entanglement entropy of charged particle pairs produced solely from the Schwinger effect. By comparing with the total entanglement entropy, we obtain the entanglement entropy produced purely from the Hawking radiation. Consequently, the two effects are distinguishable in near extremal RN black holes after the Page time. Furthermore, we also employ the brick wall model and the Pauli-Villars regularization to derive the entanglement entropy from the Schwinger effect, which gives a slightly different result with that obtained from the heat kernel method.
\end{abstract}

%% REVTEX4
%\pacs{}

\maketitle
\newpage
\tableofcontents
%%%%%%%%%%%%%%%%%%%%%%%%%%%%%%%%%%%%%%%%%%%%%%%%%%%%%%%%%%%%%%%%%%%%%%
\section{Introduction}
%%%%%%%%%%%%%%%%%%%%%%%%%%%%%%%%%%%%%%%%%%%%%%%%%%%%%%%%%%%%%%%%%%%%%%
Charged black holes can emit charged particles through spontaneous pair production which is composed of two independent mechanisms: the Hawking radiation and the Schwinger effect. It is well known that the Hawking radiation can be interpreted as a tunneling process through the black hole horizon~\cite{Parikh:1999mf}, and the Schwinger effect is the pair creation of charged particles from vacuum instability caused by strong electromagnetic field~\cite{Schwinger:1951nm}. The mechanism and properties of spontaneous pair production in charged black holes have been firstly found in the near extremal Reissner-Nordstr{\"o}m (RN) black holes~\cite{Chen:2012zn}. Subsequently, the study has been extended to various charged black holes in the asymptotically flat or (Anti)-de Sitter ((A)dS) spacetimes, see for example~\cite{Chen:2014yfa,Kim:2016dmm,Chen:2016caa,Chen:2017mnm,Chen:2020mqs,Zhang:2020apg,
Cai:2020trh,Siahaan:2023ibs,Chen:2023swn,Lin:2024jug,Siahaan:2019ysk,Chen:2025xrv}. In these cases, the near extremal limit of the charged black hole is crucial to obtain the analytical results for the pair production rate, in which the near horizon geometry contains an AdS$_2$ or a warped AdS$_3$ geometry. It was shown that the dominant contribution to the pair production rate of the RN-AdS$_5$ black hole comes from the near horizon region, by comparing with the result obtained in the full spacetime in the low frequency limit~\cite{Zhang:2020apg}. In addition, the condition for pair production in the near extremal charged black hole is consistent with the cosmic censorship conjecture~\cite{Ong:2019glf,Gan:2019ibg} and the weak gravity conjecture~\cite{Lin:2025wfe}, and the Schwinger effect can also be used to constrain the primordial black hole as dark matter~\cite{Lehmann:2019zgt,Liu:2020cds,Bai:2019zcd}. Moreover, a nontrivial property observed is that the Hawking radiation and the Schwinger pair production are intertwined and cannot be distinguished from each other in the radiation spectrum by imposing the boundary conditions~\cite{Chen:2012zn}. Therefore, it would be interesting to ask how to distinguish these two effects in the produced particle pairs.

In the present paper, we will show that the entanglement entropy is an effective probe to identify the Hawking radiation and the Schwinger effect in the charged black hole. Note that the radiation of a charged black hole (except for the extremal case) is comprised of particles both from the Hawking radiation and the Schwinger effect, whose total entanglement entropy can be calculated using the island formula given by~\cite{Penington:2019kki, Almheiri:2019qdq}
\begin{equation}\label{Island formula}
S(R)=\mathrm{min}\left \{ \mathrm{ext}\left \{ S_{\mathrm{gen}} \right \}  \right \} =\mathrm{min}\left \{ \mathrm{ext}\left \{ \frac{\mathrm{Area}(\partial I)}{4G_{N}}+S_{\rm{matter}}^{\rm{finite}}(R\cup I)  \right \}  \right \},
\end{equation}
where $I$, $R$ and $\partial I$ represent the island, the radiation region and the boundary of the island, respectively. $G_N$ is the renormalized Newtonian constant, which has absorbed the arealike divergences from the entanglement entropy of matter fields. While the contribution of island without absorbing the divergences is $\mathrm{Area}(\partial I)/(4G_B)$, where $G_B$ is the bare Newtonian constant. $S_{\mathrm{gen}}$ is the generalized entropy which includes contributions from the island and the matter field, and the entanglement entropy of the radiation is then obtained by finding the quantum extremal surface. The island formula has been extensively studied in the context of charged black holes and it was shown for non-extremal eternal charged black holes, the radiation entropy grows approximately linearly with time before the Page time and approaches roughly twice the Bekenstein-Hawking entropy after the Page time. For the extremal case, however, the entropy is ill-defined before the Page time and approximately equal to the Bekenstein-Hawking entropy after the Page time~\cite{Ling:2020laa, Wang:2021woy, Kim:2021gzd, Yu:2021cgi,Ahn:2021chg, Yadav:2022fmo, Tong:2023nvi, Lin:2024gip, Qu:2024sby}. We will apply the island formula in the near extremal charged black hole to reveal the behavior of the radiation entropy dominated by the Schwinger effect.

In order to identify the Schwinger effect from the Hawking radiation, we need to compute the entanglement entropy of charged particle pairs solely produced from the Schwinger effect. In the near horizon region of the near extremal RN black hole, the existence condition for pair production corresponds to the violation of the Breitenlohner-Freedman (BF) bound in an AdS$_2$ or a warped AdS$_3$ spacetime~\cite{Chen:2012zn}. To determine the entanglement entropy of the Schwinger mechanism, we first calculate the entropy of a massive charged scalar field in this background where the Schwinger mechanism dominates while the Hawking radiation is exponentially suppressed. The entanglement entropy of the quantum field is defined by tracing out the modes residing inside the horizon and it is given by $S=-\mathrm{Tr}\rho \ln \rho =-(\alpha \partial_\alpha -1 )\ln Z(\alpha) |_{\alpha=1}$, in which the partition function $Z(\alpha)=\mathrm{Tr}\rho^\alpha$ is the Euclidean path integral over fields defined on an $\a$-sheet covering of the smooth manifold, and the effective action $W=-\ln Z(\a)$ can be calculated from the heat kernel method~\cite{Solodukhin:1994st, Solodukhin:1995ak}. Through analyzing the transformation between the nonextremal and the near extremal RN black hole, we will verify that the entropy obtained from the heat kernel method is purely from the Schwinger effect. In addition, by introducing high curvature terms to absorb the divergences of the entropy and incorporating the contribution from the island, we obtain the entanglement entropy contributed from the Schwinger effect in the near extremal case. Subsequently, the entanglement entropy from the Hawking radiation is the difference between the total entanglement entropy and that from the Schwinger effect. Therefore, the Hawking radiation and the Schwinger effect in the pair production can be distinguished from each other by comparing the entanglement entropy. 

Moreover, we also utilize an alternative approach, i.e. the brick wall model~\cite{tHooft:1984kcu} to compute the entanglement entropy of the charged scalar field in the near extremal RN black hole, which is less rigorous but effective. The brick wall model gives the thermodynamic entropy instead of the entanglement entropy, which is one of the earliest successful methods showing that the entropy scaling with area can be naturally connected to the black hole area entropy. The method was widely applied to calculate the entropy of quantum fields in black holes~\cite{Ghosh:1994mm, Lee:1996nqa, Cognola:1997xp,Kim:1998zs, Mukohyama:1998rf, Bai:2003ud, Yoon:2007aj, Sarkar:2007uz}. Since the entropy diverges as the system approaches the horizon, the Pauli-Villars regularization scheme can be used to obtain a finite outcome~\cite{Demers:1995dq,Kim:1996nya,Solodukhin:1996jt}. We will show that, in the brick wall model, the divergence of entropy in the extremal RN black hole is non-renormalizable due to the charge of the scalar field. A possible explanation is that the brick wall model fails to fully describe the physics of extremal black holes. However, this does not have a significant impact on identifying the Schwinger effect with the Hawking radiation since only the Schwinger effect exists in the extremal limit.

The structure of the paper is as follows. In Section~\ref{Island}, we will firstly give a brief review on the near extremal RN black hole and the dynamics of a charged scalar field in the near horizon region, then we will apply the island formula to calculate the total radiation entropy of the near extremal and extremal RN black hole, respectively. In Section~\ref{heat kernel method}, we employ the heat kernel method to calculate the entropy of charged scalar field and derive the entanglement entropy purely from the Schwinger effect. Then we identify the Hawking radiation and the Schwinger effect in the pair production using entanglement entropy in Section~\ref{decouple}. In Section~\ref{brick wall}, we apply an alternative approach, i.e. the brick wall model and the Pauli-Villars regularization to compute the entropy from the Schwinger effet. Finally, conclusions and discussions are drawn in Section~\ref{con}.

%%%%%%%%%%%%%%%%%%%%%%%%%%%%%%%%%%%%%%%%%%%%%%%%%%%%%%%%%%%%%%%%%%%%%%
 \section{Island of (near) extremal RN black holes}\label{Island}
%%%%%%%%%%%%%%%%%%%%%%%%%%%%%%%%%%%%%%%%%%%%%%%%%%%%%%%%%%%%%%%%%%%%%%
 %%%%%%%%%%%%%%%%%%%%%%%%%%%%%%%%%%%%%%%%%%%%%%%%%%%%%%%%%%%%%%%%%%%%%%
\subsection{Review of near-extremal RN black holes}%\label{review}
%%%%%%%%%%%%%%%%%%%%%%%%%%%%%%%%%%%%%%%%%%%%%%%%%%%%%%%%%%%%%%%%%%%%%%
The RN black hole solution is (in units of $c = \hbar = G = 1$)
\begin{eqnarray} \label{RN}
ds^2 &=& - \left( 1 - \frac{2 M}{r} + \frac{Q^2}{r^2} \right) dt^2 + \frac{dr^2}{1 - \frac{2 M}{r} + \frac{Q^2}{r^2}} + r^2 d\Omega_2^2,
\nonumber\\
A &=& \frac{Q}{r} dt, \qquad F = \frac{Q}{r^2} dt \wedge dr,
\end{eqnarray}
where $M$ and $Q$ are respectively the mass and the charge of the black hole, $d\Omega_2^2 = d\theta^2 + \sin^2\theta d\phi^2$ represents the standard metric on a unit two-sphere. By taking the near horizon and near extremal limits, along with the time rescaling
\begin{equation} \label{rescal}
r \to Q + \epsilon \rho, \qquad M \to Q + \frac{\epsilon^2 B^2}{2 Q}, \qquad t \to \frac{\tau}{\epsilon},
\end{equation}
the near horizon near extremal RN black hole geometry is obtained
\bea \label{NHRN}
ds^2 &=& - \frac{\rho^2 - B^2}{Q^2} d\tau^2 + \frac{Q^2}{\rho^2 - B^2} d\rho^2 + Q^2 d\Omega_2^2,
\nonumber\\
A &=& -\frac{\rho}{Q} d\tau; \qquad F = \frac{1}{Q} d\tau \wedge d\rho,
\eea
in which the parameters $\epsilon\to 0$ and $B$ acts as the new horizon radius. The geometry in eq.(\ref{NHRN}) takes the form of a direct product AdS$_2 \times S^2$ with the same curvature radius $Q$.

For a massive charged probe scalar field $\Phi$ with mass $m$ and charge $q$, the action and the corresponding Klein-Gordon (KG) equation are
\bea \label{action}
&S = \int d^4x \sqrt{-g} \left( - \frac12 D^*_\alpha \Phi^* D^\alpha \Phi - \frac12 m^2 \Phi^*\Phi \right),\cr
&(\nabla_\alpha - i q A_\alpha) (\nabla^\alpha - i q A^\alpha) \Phi - m^2 \Phi = 0,
\eea
where $D_{\alpha} \equiv \nabla_{\alpha} - i q A_{\alpha}$ with $\nabla_\alpha$ is the covariant derivative. Decomposing the scalar field in spherical coordinates $\Phi(\tau, \rho, \theta, \phi) = \mathrm{e}^{-i \omega \tau + i n \phi} R(\rho) S(\theta)$, which admits a separation of variables for the KG equation, yielding
\begin{eqnarray}
\partial_\rho \left[ (\rho^2 - B^2) \partial_\rho R \right] + \left[ \frac{(q \rho - \omega Q)^2 Q^2}{\rho^2 - B^2} - m^2 Q^2 - \lambda_l \right] R &=& 0, \label{EqR}
\\
\frac1{\sin\theta} \partial_\theta (\sin\theta \partial_\theta S) - \left( \frac{n^2}{\sin^2\theta} - \lambda_l \right) S &=& 0, \label{EqS}
\end{eqnarray}
where $\lambda_l$ is a separation constant. The solution given by~(\ref{EqS}) is the standard spherical harmonics function with eigenvalue $\lambda_l = l (l + 1)$. To make the spontaneous pair production to occur, the effective mass squared must be below the Breitenlohner-Freedman (BF) bound, which requires~\cite{Chen:2012zn}
\be\label{BFb}
(m^2 - q^2) Q^2 + \left( l + \frac12 \right)^2 < 0,
\ee
namely, the inequality~(\ref{BFb}) implies that the charge of a created particle must have the charge to mass ratio $q/m>1$, which guarantees the cosmic censorship conjecture during pair production and consistent with the weak gravity conjecture.
%%%%%%%%%%%%%%%%%%%%%%%%%%%%%%%%%%%%%%%%%%%%%%%%%%%%%%%%%%%%%%%%%%%%%%
\subsection{Island in near extremal RN black holes}\label{Island ne}
Next we will calculate the total entanglement entropy of the radiation from the island formula. Note that for the near extremal RN black hole, the radiation entropy should be dominated by charged particles produced from the Schwinger effect. Besides, since we only focus on the near horizon geometry in eq.(\ref{NHRN}), our discussion is restricted to cases where both the island boundary and the radiation region boundary are near the horizon. 

From the near horizon geometry in eq.(\ref{NHRN}), the surface gravity is $\kappa=B/Q^2$. Introducing the tortoise coordinates
\be
\rho_*=\int^\rho\frac{Q^2d\rho}{\rho^2-B^2}= \left\{
\begin{array}{l}
-\frac{Q^2}{B}\mathrm{arctanh}\frac B\rho=-\frac{Q^2}{2B}\ln\frac{\rho+B}{\rho-B} \qquad \rho\geq B\,, \\
-\frac{Q^2}{B}\mathrm{arctanh}\frac \rho B=-\frac{Q^2}{2B}\ln\frac{\rho+B}{B-\rho} \qquad B\geq \rho \geq -B \,,
\end{array}
\right.
\ee
the Kruskal coordinates are
\be
U=-e^{-\kappa(t-\rho_*)}, \qquad V=e^{\kappa(t+\rho_*)}.
\ee
Then the metric eq.(\ref{NHRN}) becomes
\be\label{cfmkruskal}
ds^2=-g^2(\rho)dUdV+Q^2 d\Omega_2^2,
\ee
in which the conformal factor is
\be
g^2(\rho)=e^{-2\kappa\rho_*}\frac{Q^2(\rho^2-B^2)}{B^2}=\frac{Q^2(\rho+B)^2}{B^2}.
\ee
When the distance $L$ between the two entangling subregions is small, the finite part of entanglement entropy of the matter field can be approximated using the formula~\cite{Casini:2009}
\begin{equation}\label{matter field}
S_{\rm{matter}}^{\rm{finite}}=-\kappa_{4}c\frac{\mathrm{Area}(\partial R)}{L^{2}} ,
\end{equation}
where $\kappa_{4}$ is a constant related to the dimension and the spin and $c$ is the central charge, the distance $L$ between points $(t,\rho_a)$ and $(t,\rho_b)$ in metric eq.(\ref{cfmkruskal}) is
\be
L^2=g(\rho_a)g(\rho_b)\left[U(t, \rho_b)-U(t, \rho_a)\right]\left[V(t, \rho_a)-V(t, \rho_b)\right].
\ee
The time parts cancel out thus $L$ is spatial distance and it has different expressions for the island inside and outside the horizon. We first consider both the boundary of the island $\rho_a$ and the radiation region $\rho_b$ are outside the horizon $(\rho_b>\rho_a\geq B)$, then we have
\be
L^2=\frac{Q^2}{B^2}\left(\sqrt{(\rho_a-B)(\rho_b+B)}-\sqrt{(\rho_b-B)(\rho_a+B)}\right)^2.
\ee
The generalized entropy is obtained from eq.(\ref{Island formula}) as
\bea
S_{\rm{gen}}&=&2\, \frac{\mathrm{Area}(\partial I)}{4G_N}-2\kappa_{4}c\frac{\mathrm{Area}(\partial R)}{L^{2}}\cr
&=&\frac{2\pi(Q+\e \rho_a)^2}{G_N}-\frac{8  \pi \kappa_4 c(Q+\e \rho_b)^2 B^2}{Q^2\left(\sqrt{(\rho_a-B)(\rho_b+B)}-\sqrt{(\rho_b-B)(\rho_a+B)}\right)^2},
\eea
where the factor 2 comes from the double contributions from the left and right dual CFT~\cite{Chen:2009ht, Chen:2012zn}. The boundary of the island $\rho_a$ will be determined by extremizing the generalized entropy
\be
\der{S_{\rm{gen}}}{\rho_a}=\frac{8\pi \kappa_4 c B^2\left(\sqrt{\frac{\rho_b+B}{\rho_a-B}}-\sqrt{\frac{\rho_b-B}{\rho_a+B}}\right)}
{\left(\sqrt{(\rho_a-B)(\rho_b+B)}-\sqrt{(\rho_b-B)(\rho_a+B)}\right)^3},
\ee
which is negative when $\rho_b>\rho_a\geq B$. The generalized entropy is monotonically decreasing under this condition and attains its maximum value at $\rho_a=B$. Then we analyze the situation when the boundary of the island is inside the horizon $(\rho_b> B\geq \rho_a>-B)$, in which the generalized entropy is given by
\be\label{Sgen inside}
S_{\rm{gen}}=\frac{2\pi(Q+\e \rho_a)^2}{G_N}-\frac{8  \pi \kappa_4 c(Q+\e \rho_b)^2 B^2}{Q^2\left(\sqrt{(B-\rho_a)(\rho_b+B)}-\sqrt{(\rho_b-B)(\rho_a+B)}\right)^2}.
\ee
Taking the partial derivative with respect to $\rho_a$ gives
\be
\der{S_{\rm{gen}}}{\rho_a}=\frac{8\pi \kappa_4 c B^2\left(-\sqrt{\frac{\rho_b+B}{B-\rho_a}}-\sqrt{\frac{\rho_b-B}{\rho_a+B}}\right)}
{\left(\sqrt{(B-\rho_a)(\rho_b+B)}-\sqrt{(\rho_b-B)(\rho_a+B)}\right)^3},
\ee
which implies that the generalized entropy is increasing when $B\geq \rho_a>\frac{B^2}{\rho_b}$ while decreasing under $\frac{B^2}{\rho_b}>\rho_a>-B$, and it tends to negative infinity at the value of $\frac{B^2}{\rho_b}$, thus only reaching a peak when the boundary of the island is located at the horizon \footnote{The conclusion that the island is on the horizon can also be obtained by taking the near horizon near-extremal limit in a general RN black hole~\cite{Wang:2021woy}. }
\be\label{total ne}
S_{\rm total}=S_{\rm{gen}}(B)=\frac{2\pi Q^2}{G_N}-\frac{4 \pi c \kappa_4 B}{\rho_b-B}.
\ee
The value of the total radiation entropy at $-B$ is larger than its value at $B$. The leading term is the double Bekenstein-Hawking entropy which agrees with previous studies. Moreover, the Page curve still holds, see Appendix~\ref{Page}.

Eq.(\ref{total ne}) implies that the $B$-containing term comes from the Hawking radiation, and the first $Q$-dependent term should also contain the contribution from the Hawking radiation, for the entropy of the Hawking radiation must be positive. We will calculate this component in the following section.
%%%%%%%%%%%%%%%%%%%%%%%%%%%%%%%%%%%%%%%%%%%%%%%%%%%%%%%%%%%%%%%%%%%%%%
\subsection{Island in extremal RN black holes}\label{Island e}
In the extremal case, $B=0$, and the surface gravity $\kappa$ vanishes. The tortoise coordinate becomes
\be
\rho^*=\int^\rho\frac{Q^2 d\rho}{\rho^2}=-\frac {Q^2}{\rho}.
\ee
As in the Kruskal coordinates, the metric in eq.(\ref{NHRN}) is
\be
ds^2=-w^2(\rho)dUdV+Q^2 d\Omega_2^2, \qquad w^2(\rho) =e^{\frac{2\kappa Q^2}{\rho}}\frac{\rho^2}{\kappa^2Q^2}.
\ee
The corresponding geodesic distance between points $(t,\rho_a)$ and $(t,\rho_b)$ is
\be
L^2=\frac{\rho_a\rho_b}{\kappa^2Q^2}\left[e^{\kappa Q^2\left(-\frac{1}{\rho_a}+\frac{1}{\rho_b}\right)}+e^{\kappa Q^2\left(\frac{1}{\rho_a}-\frac{1}{\rho_b}\right)}-2\right].
\ee
Taking the $\kappa\rightarrow 0$ limit, the generalized entropy reduces to
\be
S_{\rm{gen}}=\frac{\pi(Q+\e \rho_a)^2}{G_N}-\frac{4  \pi \kappa_4 c(Q+\e \rho_b)^2 \rho_a\rho_b}{Q^2\left(\rho_a-\rho_b\right)^2}.
\ee
Then one obtains
\be
\der{S_{\rm{gen}}}{\rho_a}=\frac{4\pi c \kappa_4\rho_b(\rho_a+\rho_b)}{(\rho_a-\rho_b)^3}
\ee
which takes negative value under $\rho_b>\rho_a\geq0$. The entropy attains its extremum at $\rho_a=0$. Thus the total entanglement entropy of radiation in extremal RN black is
\be
S_{\rm{total}}=S_{\rm{gen}}(0)=\frac{\pi Q^2}{G_N},\label{Island entropy extremal}
\ee
which shows that the contribution from the matter field vanishes. The reason behind this may lies in our incomplete characterization of the entanglement entropy for matter field using eq.(\ref{matter field}). A more precise approach to calculate the entanglement entropy of matter fields is required.
%%%%%%%%%%%%%%%%%%%%%%%%%%%%%%%%%%%%%%%%%%%%%%%%%%%%%%%%%%%%%%%%%%%%%%
\section{Entanglement entropy from Schwinger effect}\label{heat kernel method}
\subsection{Heat kernel expansion}\label{near-extremal}
In this section, we will use the heat kernel method to calculate the entanglement entropy of the charged scalar field. Then we will derive the radiation entropy purely from the Schwinger effect in the (near) extremal RN black hole. The entanglement entropy of a quantum field is defined by the standard relation
\be
S=(\alpha\partial_\alpha-1)W(\alpha)|_{\alpha=1},
\label{entropy-BH}
\ee
where $W(\a)=- \ln Z(\a)$ is the one-loop effective action and $Z(\a)$ is the partition function given by the Euclidean path integral over fields defined on a Euclidean space with conical singularity. For the charged scalar field, the partition function is $Z=\mathrm{det}^{-1/2}  \mathcal{D}$, with the field operator
\be
 \mathcal{D}=-(\nabla_\alpha - i q A_\alpha) (\nabla^\alpha - i q A^\alpha) + m^2 .
\ee
Thus the effective action is expressed as
\be
W(\alpha)=-{1\over 2}\int_{\varepsilon^2}^\infty {ds\over s}\, \mathrm{Tr} K_{\alpha}(s)\, ,
\label{effective action}
\ee
and the trace of the heat kernel $K=\mathrm{e}^{-s \mathcal{D}}$ takes the small $s$ expansion as
\be
 \mathrm{Tr} K_{\alpha} (s)={1 \over (4\pi s)^2}\sum_{n=0}^{}{} {a}_n s^n\, ,\label{K expansion}
\ee
where $a_n$ takes the form ${a}_n={a}^{\rm reg}_n + a^\Sigma_{n}$. The ${a}^{\rm reg}_n$ are standard heat kernel coefficients, the first few coefficients are~\cite{Vassilevich:2003xt}
\begin{align}
&a_0^{\rm reg}=\int_{E_\alpha} 1 \ \ , \ \ a^{\rm reg}_1=\int_{E_\alpha}({1 \over 6}\bar{R}-m^2)\, ,\nonumber \\
&a^{\rm reg}_2 =\int_{E_\alpha}\left({1 \over 180} \bar{R}^2_{\mu\nu\alpha\beta} -{1 \over 180} \bar{R}^2_{\mu\nu} +{1 \over 30} \nabla^2{ \bar{R}}+{1\over 2}\left(-m^2+{1\over 6}\bar{R}\right)^2 -{1 \over 12 }q^2 F_{\mu\nu}F^{\mu\nu} \right), \label{heat kernel regular coefficients}
\end{align}
in which $E_\a$ is an $\a$-fold covering of a smooth manifold $E$ along the Killing vector $\partial_\varphi$. The coefficients due to the singular surface $\Sigma$ (the horizon) are
\begin{align}
 &a_{0}^\Sigma =0; \ \ \ a^\Sigma_{1}={\pi \over3}{(1-\alpha)(1+\alpha)\over\alpha}\int_{\Sigma}^{}1 \, , \nonumber \\
 &a^\Sigma_{2}={\pi \over 3} {(1-\alpha)(1+\alpha) \over \alpha}\int_{\Sigma}^{}({1 \over 6}\bar{R}-m^2) -{\pi \over 180}{(1-\alpha)(1+\alpha)(1+\alpha^2) \over \alpha^3}\int_{\Sigma}^{}(\bar{R}_{ii}-2\bar{R}_{ijij})\, .
 \label{heat kernel conical coefficients}
\end{align}
Here $\bar{R}_{ii}=\bar{R}_{\mu\nu}n_i^{\mu}n_i^{\nu}$, $\bar{R}_{ijij}=\bar{R}_{\mu\nu\lambda\rho}n^{\mu}_in^{\lambda}_in^{\nu}_j n^{\rho}_j $ and the surface integral $\int_\Sigma\equiv\int_\Sigma \sqrt{\gamma}d^{d-2}\theta$. $n^\mu_k \; (k=1, 2$) are two unit vectors orthogonal to the surface $\Sigma$, with $n_1^\mu=\left(\frac{Q}{\sqrt{\rho^2-B^2}},0 ,0 ,0 \right)$ and $n_2^\mu=\left(0, \frac{\sqrt{\rho^2-B^2}}{Q},0 ,0 \right)$. We present the derivation of these coefficients in Appendix \ref{heat kernel expansion}.

Then we can calculate the entropy of the charged scalar field from eq.(\ref{entropy-BH}), which gives
\be
S={A(\Sigma)\over 48\pi\varepsilon^2}-{1\over 144\pi}\int_\Sigma \left( \bar{R}-6m^2-{1\over 5}\big(\bar{R}_{ii}-2\bar{R}_{ijij}\big)\right)\ln \varepsilon, \label{UV divergence}
\ee
where $A(\Sigma)$ is the area of the horizon. Note that the result is the same as that of a neutral scalar field, which indicates that the contribution of the charge to the entanglement entropy of the scalar field is not captured by the heat kernel method (i.e., at the one-loop level). For the near extremal RN black hole
\begin{align}
\bar{R}=0,\qquad \bar{R}_{ii}=\bar{R}_{ijij}=-\frac{2}{Q^2}.
\end{align}
Consequently we obtain the entanglement entropy of the charged scalar field as
\be
S_{\rm near-extremal, \; UV}=\frac{Q^2}{12 \varepsilon^2}+\left(\frac 16 m^2 Q^2+\frac1 {90}\right)\ln \varepsilon. \label{hk scalar entropy}
\ee

\subsection{Evidence for the entropy solely originated from the Schwinger effect}\label{regular}
Note that the previous work~\cite{Chen:2012zn} used the charged scalar field to obtain the mean number of pairs which is dominated by the Schwinger effect with the Hawking radiation playing a subleading role. It seems that the entropy in eq.(\ref{hk scalar entropy}) combines the contributions from the two parts. However, eq.(\ref{hk scalar entropy}) is independent of $B$ while the Hawking temperature is $T_H=B/2\pi Q^2$, which indicates that the entropy does not contain the contribution from the Hawking radiation. This can be confirmed from analyzing the transition from nonextremal to near extremal RN black hole. By applying the same procedure for the nonextremal RN black hole metric in eq.(\ref{RN}), we obtain the entropy with contributions from both the Schwinger effect and significant Hawking radiation
\be
S_{\rm non-extremal, \; UV}=\frac{r_+^2}{12 \varepsilon^2}+\frac 16 m^2 r_+^2 \ln \varepsilon + \frac1{90}\left(\frac{3r_- -2 r_+}{r+}\right)\ln \varepsilon,\label{hk scalar entropy RN}
\ee
where $r_{\pm}=M\pm\sqrt{M^2-Q^2}$ are the outer and inner horizons. Then taking the near horizon near extremal limit in (\ref{rescal}), one obtains
\begin{align}\label{nearex scalar entropy}
S=\frac{Q^2}{12 \varepsilon^2}+\left(\frac 16 m^2 Q^2+\frac1 {90}\right)\ln \varepsilon+\e \left(\frac{BQ}{6\varepsilon^2}+\frac 13 m^2 BQ\ln \varepsilon-\frac{B}{15Q} \ln \varepsilon \right).
\end{align}
Clearly, the terms of order $\mathcal{O}(\epsilon)$ give the subleading contributions to the entropy, which are also proportional to the Hawking temperature of the near extremal RN black hole. Further taking the extremal limit $M=Q$, namely, the zero Hawking temperature limit $B=0$, the subleading terms vanish, and the result in eq.(\ref{nearex scalar entropy}) gives exactly  
that in eq.(\ref{hk scalar entropy}). Therefore, the entropy in eq.(\ref{hk scalar entropy}) is the contribution purely from the Schwinger mechanism, which shows that for the near horizon near extremal RN black hole geometry, the heat kernel method only captures the entropy from the Schwinger effect. While the island formula can capture both of the two contributions.

%%%%%%%%%%%%%%%%%%%%%%%%%%%%%%%%%%%%%%%%%%%%%%%%%%%%%%%%%%%%%%%%%%%%%%%%%%%%%%%%%%%%%%%%%%%%
\subsection{The full entropy attributable to the Schwinger effect}\label{regular}
%%%%%%%%%%%%%%%%%%%%%%%%%%%%%%%%%%%%%%%%%%%%%%%%%%%%%%%%%%%%%%%%%%%%%%%%%%%%%%%%%%%%%%%%%%%%
As shown in eq.(\ref{hk scalar entropy}), the entanglement entropy from the Schwinger effect in the near extremal black hole is UV divergent necessitating renormalization procedures. The standard procedure is to remove UV divergences by redefining the couplings that appear in the gravitational action. The bare gravitational action consists of the Einstein–Hilbert action plus all possible quadratic Riemann curvature terms
\begin{equation}
W_{gr}=\int{}\sqrt{g}d^4x \left( -{1 \over 16\pi G_B} (R  + 2\Lambda_B)
+c_{1,B}R^2+c_{2,B}R^2_{\mu \nu} +c_{3,B} R^2_{\mu\nu\alpha\beta}  - \frac 1{4g_s^2} F_{\mu\nu}^2 + \ldots \right)\, . \label{bare action}
\end{equation}
Focusing on the divergent terms in the effective action~(\ref{effective action}), we obtain
\begin{align}
W_{div}=&-{1\over 32\pi^2}\left(\frac{1}{\varepsilon^4}-\frac{m^2}{\varepsilon^2}-m^4 \ln \varepsilon \right)\int_E 1-{1\over 96\pi^2}\left(\frac{1}{2\varepsilon^2}+m^2 \ln \varepsilon \right)\int_E R\cr
&+{1\over 32\pi^2}\int_E\left({1\over 90}R^2_{\mu\nu\alpha\beta}-{1\over 90}R^2_{\mu\nu}+{1\over 36}R^2\right)\, \ln\varepsilon -{1\over 32\pi^2}\int_E \frac{q^2}{6}F_{\mu\nu}^2 \ln\varepsilon \, ,
\label{action UV divergences}
\end{align}
The renormalized coupling constants in the effective gravitational action are obtained by combining the scalar one-loop action with the original bare action
\begin{align}
W_{gr}+W_{div}=&\int_E\left\{ -\frac{1}{8\pi}\left(\frac{\Lambda_B}{G_B}+\frac{1}{4\pi \varepsilon^4}-\frac{m^2}{4\pi \varepsilon^2}-\frac{m^4}{4\pi}\ln \varepsilon \right)-\frac{1}{16\pi}\left(\frac 1 G_B + \frac{1}{12 \pi \varepsilon^2}+\frac{m^2}{6\pi}\ln \varepsilon\right)R \right. \cr
&+\left(c_{1,B}+\frac{1}{1152 \pi^2}\ln \varepsilon \right)R^2+\left(c_{2,B}-\frac{1}{2880 \pi^2}\ln \varepsilon \right)R_{\mu\nu}^2\cr
& \left. +\left(c_{3,B}+\frac{1}{2880 \pi^2}\ln \varepsilon \right)R_{\mu\nu\alpha\beta}^2 +\left(\frac 1 {4 g_s^2} - \frac{q^2}{192 \pi^2}\ln \varepsilon \right)F_{\mu\nu}^2+\ldots  \right\}.
\label{renormalized action UV}
\end{align}
These quadratic-curvature interactions will modify the radiation entropy with a term of the form
\begin{equation}
\Delta S=-\int_{\Sigma} \left( 8\pi
c_{1,B} \bar{R}+4\pi c_{2,B}\bar{R}_{ii}+8\pi  c_{3,B}\bar{R}_{ijij} \right)\, .
\label{added entropy}
\end{equation}

Since the entanglement entropy of the radiation is composed of the island part and the matter field part, for the extremal RN black hole, the entanglement entropy coming from the island is equal to the Bekenstein-Hawking entropy $\pi Q^2$ as determined in Sec~\ref{Island}, which can be identified as the entropy associated with the Schwinger mechanism. Therefore, combining all the contributions mentioned above, we obtain the full expression of entanglement entropy contributed from the Schwinger effect for the near extremal RN black hole as
\begin{align}\label{added entropy}
S_{\rm Schwinger}&= S_{\rm bare-Island}+S_{\rm near-extremal, \; UV}+\Delta S \nno\\
 &= \frac{\pi Q^2}{G_B}+ \frac{Q^2}{12 \varepsilon^2}+\left(\frac 16 m^2 Q^2+\frac1 {90}\right)\ln \varepsilon+32\pi^2 c_{2,B}+64\pi^2c_{3,B}\nno\\
&= \pi Q^2\left(\frac 1 {G_B}+\frac{1}{12 \pi \varepsilon^2}+\frac{m^2}{6\pi}\ln \varepsilon\right)+32\pi^2\left(c_{2,B}-\frac{1}{2880 \pi^2}\ln \varepsilon\right)+64\pi^2\left(c_{3,B}+\frac{1}{2880 \pi^2}\ln \varepsilon \right)\nno\\
&= \frac{\pi Q^2}{G_N}+32\pi^2 c_{2,r}+64\pi^2c_{3,r},
\end{align}
where $S_{\rm bare-Island}=\pi Q^2/G_B$ is island's contribution to the entanglement entropy without absorbing the divergences from the matter fields. Note that the entropy in eq.(\ref{added entropy}) is valid only after the Page time since island is absent before the Page time, and the derivation explicitly shows how the arealike divergences from the entanglement entropy of matter fields enters into the renormalized Newtonian constant $G_N$. After accounting for the double-sides contribution, the resulting entropy of the Schwinger effect is
\begin{equation}
S_{\rm Schwinger}=\frac{2\pi Q^2}{G_N}+64\pi^2 c_{2,r}+128\pi^2c_{3,r}.
\label{Sch_entropy_nex}
\end{equation}
Clearly, the Schwinger effect plays the leading role in the radiation. 

%%%%%%%%%%%%%%%%%%%%%%%%%%%%%%%%%%%%%%%%%%%%%%%%%%%%%%%%%%%%%%%%%%%%%%%%%%%%%%%%%%%%%%%%%%%%%%%%%%
\subsection{Entropy of Schwinger effect in the extremal RN black hole}\label{extremal heat kernel}
%%%%%%%%%%%%%%%%%%%%%%%%%%%%%%%%%%%%%%%%%%%%%%%%%%%%%%%%%%%%%%%%%%%%%%%%%%%%%%%%%%%%%%%%%%%%%%%%%%
For the extremal RN black hole, $B=0$ and the Hawking temperature vanishes so that only Schwinger effect exists. Applying the same procedure as in the near extremal case we find the entanglement entropy of the charged scalar field in the extremal RN black hole is the same as that in the near extremal case
\be
S_{\rm extremal, \; UV}=\frac{Q^2}{12 \varepsilon^2}+\left(\frac 16 m^2 Q^2+\frac1 {90}\right)\ln \varepsilon. \label{hk scalar entropy extremal}
\ee
Thus the renormalized radiation entropy of the extremal RN black hole is
\be
S_{\rm extremal}=\frac{\pi Q^2}{G_N}+32\pi^2 c_{2,r}+64\pi^2c_{3,r}.
\label{Schw_entropy_ex}
\ee
The result demonstrates that the entropy in eq.(\ref{Sch_entropy_nex}) derived in the near extremal RN black hole can be attributed to the Schwinger effect. During the transition from an extremal to a near extremal black hole, the emergence of a small Hawking temperature does not significantly change the entropy of the Schwinger effect. In the extremal case, eq.(\ref{Schw_entropy_ex}) contains an additional contribution from the matter field entropy compared to that obtained by the island formula in eq.(\ref{Island entropy extremal}). This is because the treatment of the matter field in the island formula is not enough to extract this entropy. Consequently, we adopt eq.(\ref{Schw_entropy_ex}) for the total radiation entropy of the extremal RN black hole.

%%%%%%%%%%%%%%%%%%%%%%%%%%%%%%%%%%%%%%%%%%%%%%%%%%%%%%%%%%%%%%%%%%%%%%%%
\section{Distinguish the Schwinger effect and the Hawking radiation}\label{decouple}
%%%%%%%%%%%%%%%%%%%%%%%%%%%%%%%%%%%%%%%%%%%%%%%%%%%%%%%%%%%%%%%%%%%%%%%%
By far, we have calculated the total radiation entropy in Section~\ref{Island} and the entanglement entropy from the Schwinger mechanism in Section~\ref{heat kernel method}. For near extremal black holes, the total radiation entropy receives contributions from both the Hawking radiation and the Schwinger effect. The entanglement entropy of the Hawking radiation is given by the difference between the total radiation entropy and the Schwinger-effect entropy, i.e.
\be
S_{\mathrm{Hawking}}=-64\pi^2 c_{2,r}-128 \pi^2 c_{3,r}-\frac{4 \pi c \kappa_4 B}{\rho_b-B}.
\ee

In the extremal RN black hole, the Hawking radiation entropy becomes zero because the Hawking temperature vanishes. Therefore we can distinguish the Schwinger effect from the Hawking radiation in near extremal and extremal RN black holes after the Page time (see the Tab.~\ref{table}). Before the Page time, our approach for the Schwinger-effect entropy is ineffective.
\begin{table}[!htbp]
\renewcommand{\arraystretch}{1.6}
\setlength{\tabcolsep}{18pt}
\caption{Distinguish the Schwinger effect and the Hawking radiation in near extremal RN black holes in terms of the entanglement entropy after the Page time.}\label{table}
\centering
\begin{tabular}{|c|c|c|}
\hline
                         &       Schwinger Effect                                &      Hawking Radiation                                 \\
\hline
near-extremal  &      $\frac{2 \pi Q^2}{G_N}+64\pi^2 c_{2,r}+128 \pi^2 c_{3,r}$     &    $-64\pi^2 c_{2,r}-128 \pi^2 c_{3,r}-\frac{4 \pi c \kappa_4 B}{\rho_b-B}$                     \\
\hline
extremal           &     $ \frac{\pi Q^2}{G_N}+32\pi^2 c_{2,r}+64 \pi^2 c_{3,r} $                                          &       $0$                                              \\
\hline
\end{tabular}
\end{table}

The radiation in the near extremal RN black hole is dominated by the Schwinger effect. From the viewpoint of the entanglement entropy, the Schwinger effect provides the dominant contribution to the entropy of the radiation, while the Hawking radiation is subleading.

%%%%%%%%%%%%%%%%%%%%%%%%%%%%%%%%%%%%%%%%%%%%%%%%%%%%%%%%%%%%%%%%%%%%%%%%%%%%%%%%%%%%
\section{Another effective method for the Schwinger-effect entropy}\label{brick wall}
%%%%%%%%%%%%%%%%%%%%%%%%%%%%%%%%%%%%%%%%%%%%%%%%%%%%%%%%%%%%%%%%%%%%%%%%%%%%%%%%%%%%
In this section, we will use the brick wall model and the Pauli-Villars regularization to obtain the entropy from the charged scalar field outside the (near) extremal RN black hole. Although the brick wall model gives the thermodynamic (statistical) entropy of the black hole, it is approximately equivalent to the entanglement entropy of radiation after the page time, if the unitarity is kept.
%%%%%%%%%%%%%%%%%%%%%%%%%%%%%%%%%%%%%%%%%%%%%%%%%%%%%%%%%%%%%%%%%%%%%%
\subsection{The brick wall model}\label{bri}
%%%%%%%%%%%%%%%%%%%%%%%%%%%%%%%%%%%%%%%%%%%%%%%%%%%%%%%%%%%%%%%%%%%%%%
For the near extremal RN black hole, the horizon is located at $\rho=B$ (where $B\neq0$). Under the WKB approximation $R(\rho) = e^{i S(\rho)}$, the radial equation eq.(\ref{EqR}) yields
\be\label{momentum}
p(\rho, \o, l)=\sqrt{\frac{(q \rho - \omega Q)^2 Q^2}{(\rho^2 - B^2)^2} - \frac{m^2 Q^2 +l (l + 1)}{\rho^2 - B^2} }
\ee
with $p=\der{S}{\rho}$. Following the brick wall by 't Hooft~\cite{tHooft:1984kcu} and imposing the brick wall boundary condition
\be\label{brick wall bc}
R(\rho)=0 \qquad \mathrm{for} \quad  \rho\leq B+\varepsilon \quad \mathrm{and} \quad  \rho\geq\rho_+,
\ee
where $\varepsilon$ is a small cutoff and $\rho_+$ satisfies $p(\rho_+)=0$, and the latter condition is to remove the infrared divergence. According to the semi-classical quantization rule, one has
\be \label{quantization}
\pi n(\o, l)=\int_{B+\frac{B}{15Q^2}\varepsilon^2}^{\rho_+} p(\rho, \o, l) d\rho,
\ee
in which the position of the brick wall is chosen as $\frac{B}{15Q^2}\varepsilon^2$ in order to match the result of the heat kernel method. The total number of states for a given energy $\o$ is determined by summing over the angular quantum number $l$
\bea \label{sum over l}
n(\o)&=&\frac1{\pi}\int_{B+\frac{B}{15Q^2}\varepsilon^2}^{\rho_+} d\rho \int dl (2l+1)\sqrt{\frac{(q \rho - \omega Q)^2 Q^2}{(\rho^2 - B^2)^2} - \frac{m^2 Q^2 +l (l + 1)}{\rho^2 - B^2} }.\cr
&=&\frac{2}{3\pi}\int_{B+\frac{B}{15Q^2}\varepsilon^2}^{\rho_+} d\rho \frac{\left[Q^2 \left(B^2 m^2-m^2 \rho ^2+(q \rho -Q \omega )^2\right)\right]^{3/2}}{  \left(\rho ^2-B^2\right)^2}.
\eea
The sum can be approximated by an integral where the expression under the square root is positive.  Evidently, the integrand diverges at the horizon. We  focus on the UV divergent terms and get 
\be
n(\o)=\frac{5Q^5(Q\o-qB)^3}{2B^3\pi}\frac1{\varepsilon^2}+\left(\frac{Q^3(Q\o-qB)^3}{3B^2\pi}
+\frac{Q^3q(Q\o-qB)^2}{B\pi}+\frac{Q^3m^2(Q\o-qB)}{B\pi}\right)\ln \varepsilon.
\ee
Only the integral values at the horizon are preserved, it is expected since the pair production mainly occurs near the horizon. In a thermal ensemble of charged scalar particles at fixed inverse temperature $\beta$, the free energy is given by
\be\label{free energy}
\beta F=\int d\o \dder{n(\o)}{\o} \ln \left(1-e^{-\beta(\o+q\Phi_H)}\right).
\ee
where $\beta=2\pi Q^2/B$ and $\Phi_H=-B/Q$ are respectively the inverse temperature and the chemical potential. It gives the free energy
\be
F=-\frac{\pi ^3 Q^8}{6 \beta ^4 B^3 \varepsilon^2}-\frac{\pi  m^2 Q^4 \ln \varepsilon }{6 \beta ^2 B}-\frac{2 q Q^5 \zeta (3) \ln \varepsilon }{ \pi  \beta ^3 B^2}-\frac{\pi ^3 Q^6 \ln \varepsilon}{45 \beta ^4 B^3}
\ee
And the statistical entropy of pair production can be obtained by
\be
S=\beta^2\der{F}{\beta}=\frac{2 \pi ^3 Q^6}{3 \beta ^3 B^2\varepsilon^2 }+\frac{\pi  m^2 Q^4 \ln \varepsilon}{3 \beta  B}+\frac{6 q Q^5 \zeta (3) \ln \varepsilon}{ \pi  \beta ^2 B^2}+\frac{4 \pi ^3 Q^6 \ln \varepsilon}{45 \beta ^3 B^3},
\ee
and the on-shell entropy, with $\beta=2\pi Q^2/B$, is
\be
S=\frac{Q^2}{12 \varepsilon^2 }+\frac{1}{6} m^2 Q^2 \ln \varepsilon+\frac{\ln \varepsilon}{90}+\frac{3 q Q \zeta (3) \ln \varepsilon}{2 \pi ^3}. \label{brick wall uv}
\ee
Note that the statistical entropy calculated by the brick wall model in eq.(\ref{brick wall uv}) is consistent with that obtained from the heat kernel method in eq.(\ref{hk scalar entropy}) except for the extra final term. The additional term is related to the charge of the scalar field and we demonstrate its absence in the entanglement entropy in Appendix \ref{heat kernel expansion}. The result indicates that the difference between the thermodynamic entropy and the entanglement entropy in the near extremal RN black hole lies in the charge. To remove the divergences in eq.(\ref{brick wall uv}), we will apply the Pauli-Villars regularization.
%%%%%%%%%%%%%%%%%%%%%%%%%%%%%%%%%%%%%%%%%%%%%%%%%%%%%%%%%%%%%%%%%%%%%%
\subsection{Pauli-Villars regularization}
%%%%%%%%%%%%%%%%%%%%%%%%%%%%%%%%%%%%%%%%%%%%%%%%%%%%%%%%%%%%%%%%%%%%%%
In the brick wall model, the renormalization was verified for the RN black hole using the Pauli-Villars regularization~\cite{Demers:1995dq}. From eq.(\ref{free energy}), the free energy exhibits both $\frac{1}{\varepsilon^2}$ and $\ln\varepsilon,m^2\ln\varepsilon$ divergences as $\varepsilon\to 0$. To remove the UV divergences, one can introduce three anticommuting regulator fields $\{\Phi_i,\, i=1, 2, 5\}$ with masses $\{m_1=m_2=\sqrt{\mu^2+m^2},\, m_5=\sqrt{4\mu^2+m^2}\}$ and two commuting fields $\{\Phi_3, \Phi_4\}$ with masses $\{m_3= m_4=\sqrt{3\mu^2+m^2}\}$ together with the original scalar field $\Phi=\Phi_0$ and its mass $m=m_0$. Masses of the regulator fields are dependent on the UV cutoff $\mu$. There are two constraints to be satisfied and therefore the divergent terms cancel with each other:
\be\label{constraints}
\sum_{i=0}^{5}\Delta_i=0 \quad \mathrm{and} \quad \sum_{i=0}^{5}\Delta_i m_i^2=0,
\ee
where $\Delta_0=\Delta_3=\Delta_4=1$ for the commuting fields and $\Delta_1=\Delta_2=\Delta_5=-1$ for the anticommuting fields. The total free energy is obtained by summing contributions from all individual fields as
\be\label{all contributions}
F=\sum_{i=0}^{5}\Delta_i F^i.
\ee
Thus the renormalized number of states is given by
\be\label{regulated n}
n(\o)=-\left(\frac{Q^3(Q\o-qB)^3}{3B^2\pi}+\frac{Q^3q(Q\o-qB)^2}{B\pi}\right)\sum_{i=0}^{5}\Delta_i\ln m_i^2-\left(\frac{Q^3(Q\o-qB)}{B\pi}\right)\sum_{i=0}^{5}\Delta_i m_i^2\ln m_i^2.
\ee
The renormalized free energy from eq.(\ref{free energy}) is
\be\label{regulated free energy}
F= \left(\frac{\pi ^3 Q^6}{45 \beta ^4 B^3}+\frac{2 q Q^5 \zeta (3)}{\pi  \beta ^3 B^2}\right)\mathcal{M}-\frac{\pi Q^4}{6 \beta ^2 B}\mathcal{N},
\ee
where
\bea\label{MN}
\mathcal{M}&=&-\sum_{i=0}^{5}\Delta_i\ln m_i^2=\ln\frac{(\mu^2+m^2)^2(4\mu^2+m^2)}{m^2(3\mu^2+m^2)^2}, \cr
 \mathcal{N}&=&\sum_{i=0}^{5}\Delta_i m_i^2\ln m_i^2=m^2\ln\frac{m^2(3\mu^2+m^2)^2}{(\mu^2+m^2)^2(4\mu^2+m^2)}+2\mu^2\ln\frac{(3\mu^2+m^2)^3}{(\mu^2+m^2)(4\mu^2+m^2)^2}.
\eea
Therefore, the renormalized statistical entropy is given by
\be\label{regulated entropy}
S=\frac{\pi Q^4}{3 \beta  B} \mathcal{N} -\left(\frac{4 \pi ^3 Q^6}{45 \beta ^3 B^3}+\frac{6 q Q^5 \zeta (3)}{\pi  \beta ^2 B^2}\right) \mathcal{M},
\ee
which gives the corresponding renormalized on-shell entropy of the pair production as
\be\label{on-shell entopy}
S=\frac{ Q^2}{6}\mathcal{N}-\left(\frac{1}{90}+\frac{3 q Q \zeta (3)}{2 \pi ^3}\right)\mathcal{M}=\frac{\mathcal{A}}{4}\frac{\mathcal{N}}{6\pi}-\left(\frac{1}{90}+\frac{3 q Q \zeta (3)}{2 \pi ^3}\right)\mathcal{M},
\ee
where $\mathcal{A}=4\pi Q^2$ is the area of the horizon. 

%%%%%%%%%%%%%%%%%%%%%%%%%%%%%%%%%%%%%%%%%%%%%%%%%%%%%%%%%%%%%%%%%%%
\subsection{Entropy from the Schwinger effect}
%%%%%%%%%%%%%%%%%%%%%%%%%%%%%%%%%%%%%%%%%%%%%%%%%%%%%%%%%%%%%%%%%%%
The entropy can be also verified as the pure Schwinger-effect enropy when considering the transformation between the nonextremal and near extremal RN black hole. The procedure is similar and will not be repeated here. 

Further taking the contribution of the island and the higher-order curvature corrections into account, the Schwinger-effect entropy is given by 
\begin{align}
S=\ &\frac{\pi Q^2}{G_B}+\frac{ Q^2}{6}\mathcal{N}-\left(\frac{1}{90}+\frac{3 q Q \zeta (3)}{2 \pi ^3}\right)\mathcal{M}+32\pi^2 c_{2,B}+64\pi^2c_{3,B}\cr
=\ &\pi Q^2\left(\frac 1 {G_B}+\frac{ Q^2}{6}\mathcal{N}\right)+32\pi^2\left(c_{2,B}-\frac{1}{2880 \pi^2}\mathcal{M} \right)+64\pi^2\left(c_{3,B}+\frac{1}{2880 \pi^2}\mathcal{M} \right)-\frac{3 q Q \zeta (3)}{2 \pi ^3}\mathcal{M}\cr
=\ &\frac{\pi Q^2}{G_N}+32\pi^2 c_{2,r}+64\pi^2c_{3,r}-\frac{3 q Q \zeta (3)}{2 \pi ^3}\mathcal{M}.
\label{added entropy in brick wall}
\end{align}
The result shares a similar form with that given by the heat kernel method in eq.(\ref{Sch_entropy_nex}). An analogous calculation for neutral scalar field was presented in~\cite{Demers:1995dq}. Apart from the couplings, the thermodynamic entropy of the radiation differs from its entanglement entropy by a contribution related to the scalar field charge.

%%%%%%%%%%%%%%%%%%%%%%%%%%%%%%%%%%%%%%%%%%%%%%%%%%%%%%%%%%%%%%%%%%%%%%
\subsection{The entropy in extremal RN black hole}\label{extremal}
%%%%%%%%%%%%%%%%%%%%%%%%%%%%%%%%%%%%%%%%%%%%%%%%%%%%%%%%%%%%%%%%%%%%%%
In the extremal RN black hole, though the entropy in eq.(\ref{on-shell entopy}) is independent on $B$, it does not mean that the entropy in extremal black holes is the same as that in the near extremal case, actually, they are different~\cite{Carroll:2009maa}. Explicitly, from the brick wall model, setting $B=0$ in eq.(\ref{sum over l}), we can obtain the number of states
\be
n(\o)=\frac{2 Q^6 \omega ^3}{9 \pi  \varepsilon^3}-\frac{q Q^5 \omega ^2}{\pi  \varepsilon^2}-\frac{Q^4 \omega  \left(m^2-2 q^2\right)}{\pi  \varepsilon}+\frac{q Q^3 \left(2 q^2-3 m^2\right)}{6 \pi } \ln (\varepsilon^2).
\ee
The free energy is
\be
F=-\frac{2 \pi ^3 Q^6}{135 \beta ^4 \varepsilon ^3}+\frac{2 q Q^5 \zeta (3)}{\pi  \beta ^3 \varepsilon ^2}+\frac{\pi  Q^4 \left(m^2-2 q^2\right)}{6 \beta ^2 \varepsilon }-\frac{q Q^3  \left(2 q^2-3 m^2\right) \zeta(1)}{6 \pi  \beta }\ln \left(\varepsilon ^2\right).
\ee
Then the statistical entropy in the extremal RN black hole is
\be
S=\frac{8 \pi ^3 Q^6}{135 \beta ^3 \varepsilon ^3}-\frac{6 q Q^5 \zeta (3)}{\pi  \beta ^2 \varepsilon ^2}-\frac{\pi  Q^4 \left(m^2-2 q^2\right)}{3 \beta  \varepsilon } +\frac{q Q^3  \left(2 q^2-3 m^2\right)\zeta(1) }{6 \pi }\ln \left(\varepsilon ^2\right).
 \ee
After using the Pauli-Villars regularization
\be
n(\o)=-\frac{qQ^3}{2\pi}\sum_{i=0}^{5}\Delta_i m_i^2\ln m_i^2+\frac{q^3Q^3}{3\pi}\sum_{i=0}^{5}\Delta_i \ln m_i^2.
\ee
The corresponding free energy gives
\be
F=\frac{qQ^3\zeta(1)}{2\pi\beta}\mathcal{N}+\frac{q^3Q^3\zeta(1)}{3\pi\beta}\mathcal{M},
\ee
in which $\mathcal{M}$ and $\mathcal{N}$ are the same those in eq.(\ref{MN}). The resulting renormalized statistical entropy is
\be\label{entropy in ebl}
S=-\frac{qQ^3\zeta(1)}{2\pi}\mathcal{N}-\frac{q^3Q^3\zeta(1)}{3\pi}\mathcal{M}.
\ee
Note that $\zeta(1)=\sum_{n=1}^{\infty}\frac1n$ diverges, causing the entropy in eq.(\ref{entropy in ebl}) in the extremal RN black hole divergent. Eq.(\ref{entropy in ebl}) differs from the result in~\cite{Demers:1995dq, Ghosh:1994mm} where the scalar field entropy vanishes for extremal black holes. The divergence occurs because the produced particles are charged, the entropy vanishes when $q=0$. However, a divergent entropy is not physical, the reason may be that brick wall model cannot capture the correct information in extremal black holes. While the result for the entanglement entropy computed from the heat kernel method suggests that this component makes no contribution, which, to some extent, explains this issue.

%%%%%%%%%%%%%%%%%%%%%%%%%%%%%%%%%%%%%%%%%%%%%%%%%%%%%%%%%%%%%%%%%%%%%%
\section{Conclusion and discussion}\label{con}
%%%%%%%%%%%%%%%%%%%%%%%%%%%%%%%%%%%%%%%%%%%%%%%%%%%%%%%%%%%%%%%%%%%%%%
In this paper, we studied the entanglement entropy of charged particle pairs produced from the Schwinger effect and the Hawking radiation in the (near) extremal RN black hole, aimed to distinguish these two mechanisms in the radiation spectrum. Our strategy was to calculate the total entanglement entropy of radiation by using the island formula, and compute the entanglement entropy contributed purely from the Schwinger effect using the heat kernel method and the brick wall method, then the entanglement entropy that comes from the Hawking radiation is the difference between them. While in the extremal RN black hole, such a distinction becomes unnecessary as only the Schwinger effect exists. Our results showed that the entanglement entropy can provide an effective tool to identify the contributions from the two mechanisms. 

The distinction between the Schwinger effect and the Hawking radiation is far from straightforward in a general black hole background in eq.(\ref{RN}), in which the two effects exhibit significant mixing. The distinction cannot be made solely based on parameter dependence. Namely, the entropy of the Schwinger effect may depend on the black hole mass $M$ and the entropy of the Hawking radiation might involve the charge $Q$. From the perspective of the Schwinger effect, virtual pairs can be separated into real particles by not only strong electromagnetic fields but also tidal forces~\cite{Wondrak:2023zdi}. The spacetime curvature can take the role of the electric field strength. As a result, the Schwinger effect could no longer be merely related to the electric field. While from the standpoint of the Hawking radiation, the violation of BF bound requires that particles produced via the Hawking radiation are charged. Thus the Hawking radiation can also couple to electromagnetic fields. The coupling makes it difficult to separate the two effects in nonextremal black holes.

There remain several aspects worthy of further exploration. Firstly, it would be interesting to study the separation of the two effects in a charged rotating black hole or in the asymptotically (A)dS spacetime. Secondly, in this work we only considered the entropy of the pair production of scalar particles. This framework can be generalized to spinor pair production. Finally, when $M\simeq Q$, the near horizon geometry of~(\ref{RN}) reduces to AdS$_2\times S_2$, in which the radiation is dominated by the Schwinger effect. In contrast, for $M\gg Q$, the geometry is approximately Rindler$_2 \times S_2$~\cite{Kim:2007ep}, where the Hawking radiation takes the leading role. The two effects could be decoupled in the Rindler$_2 \times S_2$ spacetime in a similar way.

%%%%%%%%%%%%%%%%%%%%%%%%%%%%%%%%%%%%%%%%%%%%%%%%%%%%%%%%%%%%%%%%%%%%%%
\section*{Acknowledgement}
%%%%%%%%%%%%%%%%%%%%%%%%%%%%%%%%%%%%%%%%%%%%%%%%%%%%%%%%%%%%%%%%%%%%%%
We would like to thank Chiang-Mei Chen and Sang-Pyo Kim for useful discussions. This work was supported by the National Natural Science Foundation of China (No.~12475069) and Guangdong Basic and Applied Basic Research Foundation (No.~2025A1515011321).

%%%%%%%%%%%%%%%%%%%%%%%%%%%%%%%%%%%%%%%%%%%%%%%%%%%%%%%%%%%%%%%%%%%%%%
\begin{appendix}
%%%%%%%%%%%%%%%%%%%%%%%%%%%%%%%%%%%%%%%%%%%%%%%%%%%%%%%%%%%%%%%%%%%%%%
\section{Page curve and Page time}\label{Page}
Before the Page time, the near horizon metric alone cannot completely describe the radiation entropy. In order to reproduce the Page curve, one must derive the entropy in the general metric~(\ref{RN}) and then apply the coordinates transformation. We directly use the results of the previous work~\cite{Wang:2021woy}. In the near-extremal black hole, when the island contribution is absent, the radiation entropy is purely determined by the entropy from the matter field and increases linearly with time.
\be
S\simeq \frac c3 \kappa_+ t.
\ee
Here $\kappa_+=\frac{r_+-r_-}{2r_+^2}$. Taking the near horizon near-extremal limits~(\ref{rescal}), the entropy yields
\be\label{without island}
S\simeq \frac{cB\tau}{3Q^2}.
\ee
By comparing the entropy without island~(\ref{total ne}) and with island~(\ref{without island}), one has the Page time
\be
\tau_{\mathrm{Page}}\simeq \frac{6\pi Q^4}{c B}.
\ee

In the case of the extremal black hole, the radiation entropy is divergent~\cite{Kim:2021gzd}, rendering both the entropy itself and the Page time ill-defined.
\begin{table}[!htbp]
\renewcommand{\arraystretch}{1.3}
\setlength{\tabcolsep}{15pt}
\caption{Entanglement entropy and Page time in near-extremal and extremal RN black holes}
\centering
\begin{tabular}{|c|c|c|c|}
\hline
RN Black holes             &       without island                                       &      with island                              &           Page time                                                                         \\
\hline
near-extremal                &      $S_R \simeq \frac{cB\tau}{3Q^2}$     &    $S_R \simeq 2S_{BH}$           &    $\tau_{\mathrm{Page}}\simeq\frac{6\pi Q^4}{c B}$       \\
\hline
extremal                        &     ill-defined                                                &       $S_R \simeq S_{BH} $          &          ill-defined                                                                        \\
\hline
\end{tabular}
\end{table}
%%%%%%%%%%%%%%%%%%%%%%%%%%%%%%%%%%%%%%%%%%%%%%%%%%%%%%%%%%%%%%%%%%%%%%
\section{The derivation of the heat kernel expansion of operator $\mathcal{D}=-(\nabla_\alpha - i q A_\alpha) (\nabla^\alpha - i q A^\alpha) + m^2$}\label{heat kernel expansion}
Now we will derive~(\ref{heat kernel regular coefficients}) and~(\ref{heat kernel conical coefficients}). We mostly follow~\cite{Vassilevich:2003xt}. The regular heat kernel coefficients are 
\begin{align}
a_{0}^{\mathrm{reg}}&=\int_{E_{\a}} 1, \qquad a_1^{\mathrm{reg}}=\frac 16 \int_{E_{\a}}\left(6P+\bar{ R} \right), \cr
   a_2^{\mathrm{reg}}&=\frac{1}{360}\int_{E_{\a}}\left(60\nabla^2P  +60\bar{ R} P+180P^2 +12\nabla^2\bar{ R}+5\bar{ R}^2-2 \bar{ R}_{\mu\nu}^2+2 \bar{ R}_{\mu\nu\a\beta}^2+30\Omega_{\mu\nu}^2\right).
\end{align}
where the endomorphism $P=-m^2$ and the field strength $\Omega_{\mu\nu}=-iqF_{\mu\nu}$ for the operator $\mathcal{D}$ is easily to get. Therefore the coefficients are 
\begin{align}
a_{0}^{\mathrm{reg}}&=\int_{E_{\a}} 1,  \qquad a_1^{\mathrm{reg}}=\int_{E_{\a}}\left(-m^2 + \frac 16 \bar{R}\right), \cr
   a_2^{\mathrm{reg}}&=\int_{E_{\a}}\left(\frac{1}{180}\bar{R}_{\mu\nu\a\beta}^2-\frac{1}{180}\bar{R}_{\mu\nu}^2+\frac{1}{30}\nabla^2 \bar{R}+\frac 12 \left(-m^2+\frac 16 \bar{R} \right)^2-\frac 1 {12} q^2F_{\mu\nu}F^{\mu\nu}\right).
\end{align}

Then we can deduce the coefficients due to the singular surface $\Sigma$. The heat kernel defined on the smooth manifold $E$ in powers of the proper-time parameter $s$ is
\begin{equation}
K_{E}(x,x',s)|_{s\rightarrow 0}={ e^{-\sigma ^2(x,x') /4s} \over (4\pi s)^{2}} \triangle^{1/2}(x,x')\sum_{n=0}^{\infty}a_n(x,x')s^n,
\label{eq:smex}
\end{equation}
where $\sigma(x,x')$ is the geodesic distance between points $x$, $x'$.  $\triangle(x,x')$ is the Van Vleck determinant
\begin{equation}
\triangle(x,x')=-[g(x)g(x')]^{-1/2}\det\left({\partial ^2
\sigma(x,x')/2 \over \partial x^{\mu} \partial x'^{\nu}}\right)~~~.
\label{eq:biscalar}
\end{equation}
Now the $\Sigma$ is the fix-point set of $O(2)$ isometry group. About a point $p_0$ on $\Sigma$, define Riemann normal coordinates $x^{\mu}$. We use a rescaled normal coordiantes $y^{\mu}=s^{-1/2}x^{\mu}$. Then (\ref{eq:smex}) can be expanded as
\begin{equation}
K_{E}|_{p_0}={e^{-(y-y')^2/4} \over (4\pi s)^2}
\left(1+\sum_{n=0}^{\infty}b_{(n+2)/2}(y,y',p_0)s^{(n+2)/2}\right)
\label{eq:exx0}
\end{equation}
With the known asymptotic formulas
\begin{equation}
\triangle ^{1/2}=1+{s \over 12}\bar{R}_{\mu\nu}(p_0)
(y-y')^{\mu}(y-y')^{\nu}+O(s^{3/2})~~~,
\label{eq:as1}
\end{equation}
\begin{equation}
\sigma ^2=s\left((y-y')^2-\frac s3 \bar{R}_{\mu\lambda\nu
\rho}(p_0)(y-y')^{\mu}(y-y')^{\nu}y^{\lambda}y^{\rho}+
O(s^{3/2})\right)~~~,
\label{eq:as2}
\end{equation}
we can obtain $b_1$:
\begin{equation}
b_1(y,y',p_0)=a_1+{1 \over 12}\left(\bar{R}_{\mu\nu}(p_0)(y-y')^{\mu}
(y-y')^{\nu}+\bar{R}_{\mu\lambda\nu\rho}(p_0)(y-y')^{\mu}(y-y')^{\nu}y^{\lambda}
y^{\rho}\right)
\label{eq:b0}
\end{equation}
With $a_1=\frac 16 \bar{R} -m^2$ in our background. We use the coordinates $y^\mu=(u, \varphi, v^\beta)$. The first two coordinates are orthogonal to $\Sigma$. Consider a point $p$ being close enough to $\Sigma$, and the line from $p$ to $p_0$ is orthogonal to $\Sigma$. The rescaled Riemann coordinates is $y=(u\cos \varphi, u\sin \varphi, v^\beta=0)$, for simplicity, we dismiss the coordinates $v^\beta=0$. One has the following relation:
\begin{align}
&y^i=(u\cos \varphi, u\sin \varphi),\qquad y'^i=(u\cos\left (\varphi+\o\right), u\sin\left (\varphi+\o \right)),\cr
 & Y^i=y^i-y'^i =2u\sin \frac{\o}{2}(\sin\left(\frac \o 2\right), -\cos \left(\frac \o 2\right) ),\cr
 &|y|^2=u^2, \qquad |Y|^2=4 u^2 \sin^2\left(\frac \o 2\right) , \qquad y \, \cdot Y =2 u^2 \sin^2\left(\frac \o 2\right).
\end{align}
Thus  (\ref{eq:exx0})  and (\ref{eq:b0}) yields
\begin{align}
K_{E}&={e^{-u^2\sin^2(w/2)} \over (4 \pi s)^{2}}\sum_{n=0}^{\infty}b_{n}(u^2,w,\theta)s^n \cr
b_1(u^2,w )&=a_1+\frac 16 u^2 \sin^2\left(\frac \o 2\right)\bar{R}_{ii}+{1 \over 6}u^4\left(\sin^2\left(\frac \o 2\right)-\sin^4\left(\frac \o 2\right)\right)\bar{R}_{ijij}.
\end{align}
Notice all half integer powers of $s$ in  (\ref{eq:exx0}) disappear due to the isometry. The heat kernel on the manifolds with conical singularities obeys the Sommerfeld formula
\begin{equation}
K_{E_\alpha}- K_{E}  = \frac{ i}{4\pi\alpha}\int_\Gamma \cot \frac{w}{2\alpha} K_{E} dw \,.
\label{Sommerfeld}
\end{equation}
The contour $\Gamma$ is defined by two vertical lines: one extending from $-\pi + i\infty$ to $-\pi - i\infty$, and the other from $\pi - i\infty$ to $\pi + i\infty$. It can be associated with the coefficients by taking the trace
\begin{equation}
\mathrm{Tr} \left( K_{E_\alpha}- K_{E} \right) = \int_{\Sigma_\e}\left( K_{E_\alpha}- K_{E} \right)  = {1 \over (4\pi s)^{2}}\sum_{n=0}^{\infty}a_{n+1}^\Sigma s^{n+1} \,.\label{trace}
\end{equation}
The integral over $\Sigma_{\epsilon}$ admits a representation via curvature corrections to the volume of the product space $C_{\alpha}^{r_\epsilon} \times \Sigma$, with $r_\epsilon$ interpreted as the cone radius.
\begin{equation}
\int_{\Sigma_{\epsilon}}=
\int_{0}^{2\pi\alpha}d\varphi \int_{0}^{r_\epsilon /\sqrt{s}}udu~s
\sum_{n=0}^{\infty}d_n u^{2n}s^n\int_{\Sigma}~~~,
\label{eq:measure}
\end{equation}
with
\begin{equation}
d_0=1~~~,~~~d_1={1 \over 6}\bar{R}_{ijij} -\frac 14 \bar{R}_{ii}~~~.
\label{eq:meascoef}
\end{equation}
Thus (\ref{trace}) becomes
\begin{align}
&\frac{1 }{ (4 \pi s)^2}\sum_{n=0}^{\infty}\left(\int_{\Sigma}\frac i2 \int_{\Gamma}\cot(\frac{\omega}{2 \alpha})dw\int_{0}^{r_\epsilon/\sqrt{s}} udu e^{-u^2\sin^2\left(\frac \o 2\right)}\sum_{m=0}^{n}d_mb_{n-m}(u^2,w)u^{2m}\right)s^{n+1}\cr
= &{1 \over (4\pi s)^{2}}\sum_{n=0}^{\infty}a_{n+1}^\Sigma s^{n+1}  .
\end{align}
We can read off the form of $a_1^{\Sigma}$ :
\begin{align}
a_1^{\Sigma}&=\int_{\Sigma}\frac i2 \int_{\Gamma}\cot(\frac{\omega}{2 \alpha})dw\int_{0}^{r_\epsilon/\sqrt{s}} udu e^{-u^2\sin^2\left(\frac \o 2\right)} d_0 b_0=\int_{\Sigma}\frac{\pi}{3}\frac{(1-\a)(1+\a)}{\a},
\end{align}
and $a_2^{\Sigma}$
\begin{align}
a_2^{\Sigma}=&\int_{\Sigma}\frac i2 \int_{\Gamma}\cot(\frac{\omega}{2 \alpha})dw\int_{0}^{r_\epsilon/\sqrt{s}} udu e^{-u^2\sin^2\left(\frac \o 2\right)}\left( d_0 b_1+  d_1 b_0 u^2 \right)\cr
=&{\pi \over 3} {(1-\alpha)(1+\alpha) \over \alpha}\int_{\Sigma}^{}({1 \over 6}\bar{R}-m^2) -{\pi \over 180}{(1-\alpha)(1+\alpha)(1+\alpha^2) \over \alpha^3}\int_{\Sigma}^{}(\bar{R}_{ii}-2\bar{R}_{ijij})\, . \label{derivation of a2sigma}
\end{align}
In the derivation of $a_2^{\Sigma}$, we only use terms related to $a_1^{\mathrm{reg}}$ which receives no correction from the charged scalar field. Thus (\ref{derivation of a2sigma}) is the same as the neutral scalar field.

\end{appendix}
%%%%%%%%%%%%%%%%%%%%%%%%%%%%%%%%%%%%%%%%%%%%%%%%%%%%%%%%%%%%%%%%%%%%%%

%%%%%%%%%%%%%%%%%%%%%%%%%%%%%%%%%%%%%%%%%%%%%%%%%%%%%%%%%%%%%%%%%%%%%%

\end{document}